\documentclass[aps,prl,reprint,superscriptaddress,showpacs]{revtex4-1}
\usepackage{graphicx}
\usepackage{amsmath}

\newcommand{\mr}[1]{\mathrm{#1}}
\newcommand{\vect}[1]{\boldsymbol{#1}}
\newcommand{\vectop}[1]{\widehat{\boldsymbol{#1}}}
\newcommand{\ii}{\mathrm{i}}
\newcommand{\muvec}{\boldsymbol{\mu}}
\newcommand{\gmatrix}{\mathbf{g}}
\newcommand{\ggT}{\gmatrix\gmatrix^\text{T}}
\newcommand{\normsquared}[1]{\left|#1\right|^2}

\begin{document}
\title{General Magnetic Transition Dipole Moments for Electron Paramagnetic Resonance}
\author{Joscha Nehrkorn}
\email{joscha.nehrkorn@helmholtz-berlin.de}
\affiliation{Berlin Joint EPR Laboratory, Institut f\"ur Silizium-Photovoltaik, Helmholtz-Zentrum Berlin f\"ur Materialen und Energie, Berlin, Germany}
\author{Alexander Schnegg}
\affiliation{Berlin Joint EPR Laboratory, Institut f\"ur Silizium-Photovoltaik, Helmholtz-Zentrum Berlin f\"ur Materialen und Energie, Berlin, Germany}
\author{Karsten Holldack}
\affiliation{Institut f\"ur Methoden und Instrumentierung der Forschung mit Synchrotronstrahlung, Helmholtz-Zentrum Berlin f\"ur Materialen und Energie, Berlin, Germany}
\author{Stefan Stoll}
\email{stst@uw.edu}
\affiliation{Department of Chemistry, University of Washington, Seattle, USA}
\date{\today}

\begin{abstract}
We present general expressions for the magnetic transition rates in beam Electron Paramagnetic Resonance (EPR) experiments of anisotropic spin systems in the solid state. The expressions apply to general spin centers and arbitrary excitation geometry (Voigt, Faraday, and intermediate). They work for linear and circular polarized as well as unpolarized excitation, and for crystals and powders. The expressions are based on the concept of the (complex) magnetic transition dipole moment vector. Using the new theory, we determine the parities of ground and excited spin states of high-spin (${S = 5/2}$) Fe$^\text{III}$ in hemin from the polarization dependence of experimental ground state EPR line intensities.
\end{abstract}


\maketitle

Electron paramagnetic resonance (EPR) is a spectroscopic technique that yields unique information on structural \cite{Alexander14, Klare13, Jeschke02, Duss14, Borbat01,Fehr14,Lubitz07}, magnetic \cite{Caneschi91,Maeda08,Takahashi11} and electronic properties \cite{Stutzmann85, Umeda96, George13} of paramagnetic states in material systems ranging from proteins to nanomagnets and semiconductors. In addition, EPR methods are increasingly used for controlled manipulation of spin systems, which may form the basis of spin quantum computing \cite{Childress06, Takahashi09, McCamey10, Bertaina09}.  Experimental design, interpretation, prediction and control of the latter require general theoretical tools to calculate EPR transition energies and probabilities. These properties are determined by the spin center under study and the choice of the experimental parameters.

In a standard EPR experiment, linearly polarized low-frequency microwave (mw) radiation is coupled into a resonator exposed to a static magnetic field $\vect{B}_0$ such that the radiation magnetic field component $\vect{B}_1(t)$ of the ensuing standing wave is perpendicular to $\vect{B}_0$. With a linear detector, the measured EPR spectral intensity is proportional to the power absorbed by the sample, which in turn is proportional to the quantum mechanical transition rate. For this standard geometry, compact expressions for the EPR transition rate can be found in the literature \cite{Lund04, Stoll06}. Analytical expressions for a single spin without fine or hyperfine interactions are known \cite{Bleaney60, Kneubuehl61, Pilbrow69, Rockenbauer74, Iwasaki74, Rockenbauer76}.

However, the limitation to a resonator, linear mw polarization and orthogonal orientation between static and oscillating magnetic fields restricts the versatility of EPR experiments. Recently, experimental setups that go beyond these limitations have become more prevalent. Novel non-resonant beam EPR setups explore very broad field (up to 30 T) and frequency (up to THz) ranges. These high field/high frequency EPR experiments are based on a range of excitation sources, ranging from lab-based semiconductors, lasers and tube sources \cite{Disselhorst95, Hassan00, vanSlageren03, vanTol05} to synchrotrons \cite{Talbayev04, Mihaly04, Brion07, Schnegg09} and free electron lasers \cite{Zvyagin09, Takahashi12}. Despite the variety of source technology, these approaches are all based on quasioptical techniques that transmit mw or THz radiation in open space instead of wave guides or coaxial cables. This provides much larger freedom for the alignment of the radiation beam relative to the external magnetic field.

Thereby, entirely new EPR experiments became possible. These include experiments in Faraday geometry \cite{vanSlageren05}, and the employment of split ring resonator arrays as THz metamaterials for selective EPR excitation \cite{Schneider09, Scalari12}. Unlike with resonators, in non-resonant setups circular or unpolarized radiation can be employed. Circularly polarized radiation can be used to determine the sign of g factors \cite{Hutchison60, Chang64, Suematsu74, Rockenbauer76, Henderson08} and is a possible selection tool in EPR based quantum computing \cite{Leuenberger01}. Depending on the handedness of the circularly polarized radiation, different sets of EPR transitions can be adressed in single-molecule magnets \cite{vanSlageren05}. For dynamic nuclear polarization, it was recently shown that the enhancement depends on the handedness of the circular polarized mw radiation \cite{Armstrong10}. Unpolarized radiation, or radiation that is extracted from beam paths which do not conserve the polarization of the radiation, are used in high-field cw EPR \cite{Barra90, Tarasov91, Dorlet00, Hassan00, vanTol05, Golze06}, for frequency-swept cw EPR \cite{vanSlageren04, Schnegg09, Nehrkorn13} as well as for free-electron laser based cw EPR experiments \cite{Zvyagin09}.

Spectral intensities from these new experimental designs cannot be described by current theory, which is limited to perpendicular and parallel excitation geometries with linear polarization. Here, we derive compact and general expressions for EPR magnetic transition intensities that cover all excitation geometries and polarizations. We show that the transition intensities can be described in an elegant way using a general magnetic transition dipole moment (mtdm) vector $\muvec$. The mtdm is the magnetic analog of the electric transition dipole moment vector widely used in optical spectroscopy.

First, we treat a solid-state sample containing isolated identically oriented spin centers, each containing $N$ coupled spins (electrons and/or nuclei) with arbitrary anisotropic interactions. We then extend the treatment to dilute powder samples, where isolated spin centers occur in a random uniform orientational distribution. Next, we treat the special case of isolated electron spins without fine or hyperfine interactions. All cases cover linear, circular, and unpolarized radiation. The associated derivations are given in the supplemental material (SM). Finally, we show data from an experiment that directly determines the parities of magnetic states involved in an EPR transition. This illustrates the utility of the newly derived theory.

\begin{figure}
  \includegraphics[width=85mm]{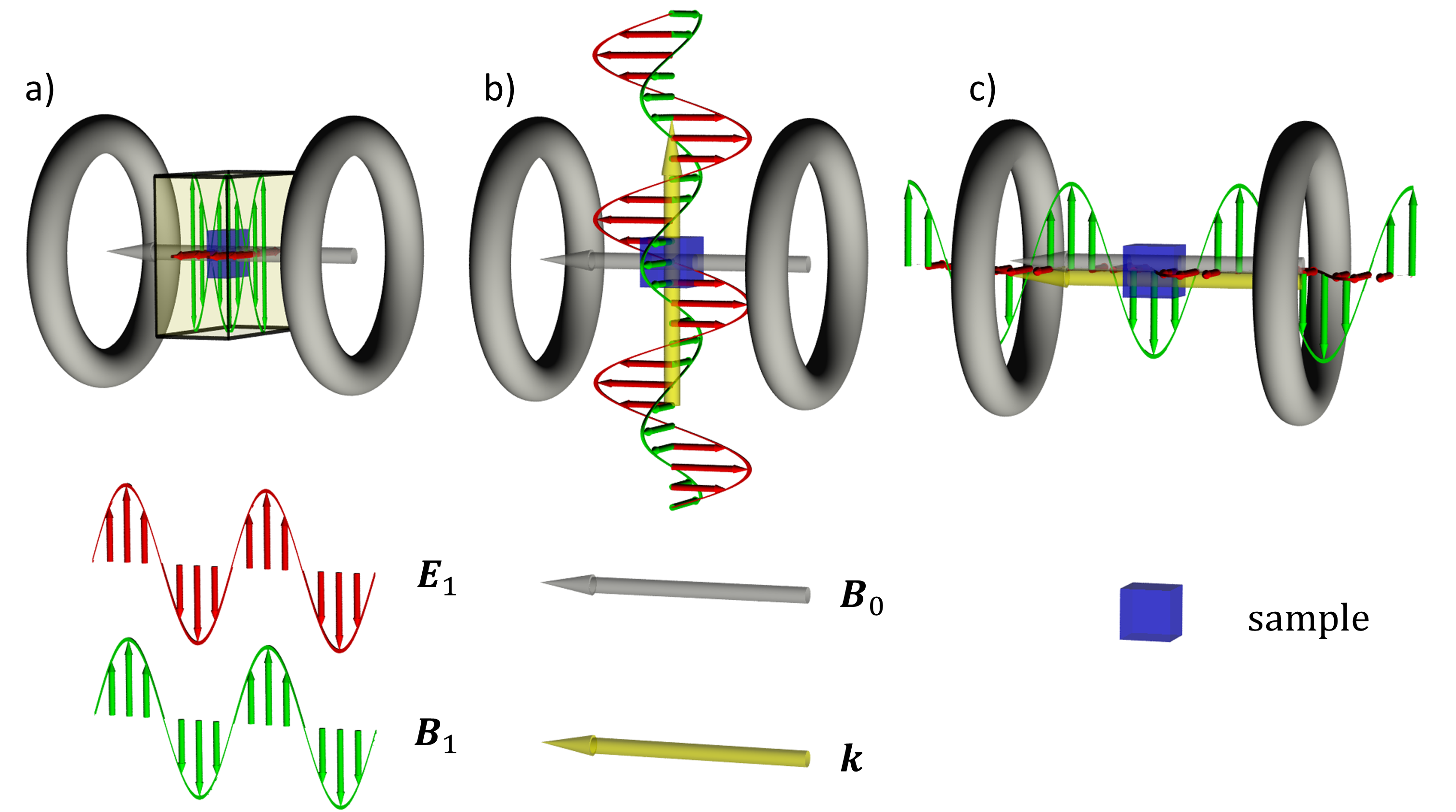}%
  \caption{Sketches of magneto-optical excitation geometries described in the text. Electric ($\vect{E}_1$) and magnetic ($\vect{B}_1$) field components of the mw/THz radiation are depicted by red and green oscillating lines, respectively.  $\vect{k}$ (yellow arrow) denotes the propagation direction of the radiation. The static magnetic field $\vect{B}_0$ is indicated by the gray arrow. Magnet coils are shown as gray tori. 
  (a) EPR excitation of a sample (blue box) inside a mw resonator (pale yellow box with black ceiling). The standing wave in the resonator ensures maximum $\vect{B}_1$  and minimum $\vect{E}_1$ at the sample position. In the present case $\vect{B}_1$ is aligned perpendicular to $\vect{B}_0$. (b) and (c) depict travelling wave excitation in Voigt geometry (b), where ${\vect{k}\!\perp\!\vect{B}_0}$, and Faraday geometry (c), where ${\vect{k}\|\vect{B}_0}$.
  \label{fig:ex_shemes}}
\end{figure}

An EPR transition between two stationary states $|a\rangle$ and $|b\rangle$ of an isolated spin center is induced by the resonant interaction between the total magnetic dipole moment of the spins and the magnetic field component $\vect{B}_1$ of the mw radiation. Since the mw radiation is always a weak perturbation to the spin center, the cw EPR line intensity  for the transition $|a\rangle\rightarrow|b\rangle$ is accurately described by time-dependent perturbation theory (Fermi's Golden Rule \cite{Cohen-Tannoudji}) and is given by
\begin{equation}\label{eq:fermi}
I_{ab} = 
\chi \frac{B_1^2}{4}
\left| \vect{n}_1^\text{T}\langle b | \widehat{\muvec} | a \rangle\right| ^2
=
\chi \frac{B_1^2}{4}
\left| \vect{n}_1^\text{T}\muvec\right|^2
=
\chi \frac{B_1^2}{4}D.
\end{equation}
$\chi$ contains the population difference of the two states, and, for field-swept spectra, the additional factor ${(\mathrm{d}(E_b-E_a)/\mathrm{d}B_0)^{-1}}$ , which is the general form of the Aasa-V\"{a}nng\aa rd 1/g factor \cite{Aasa75,vanVeen78}. $B_1$ is the (maximal) amplitude of the oscillatory $\vect{B}_1$, and $\vect{n}_1$ is a vector describing its direction. $\vectop{\mu}$ is the total magnetic dipole moment operator. For an isolated spin system with $N$ coupled spins, it is given by  ${\vectop{\mu} = \sum_{q=1}^N \sigma_q \mu_q \gmatrix_{q}\vectop{S}_{q}}$. For electron spins, ${\sigma_q = -1}$ and $\mu_q$ is the Bohr magneton $\mu_\text{B}$. For nuclear spins, ${\sigma_q = +1}$ and $\mu_q$ is the nuclear magneton $\mu_\text{N}$. $\gmatrix_q$ is the 3x3 g matrix and $\vectop{S}_{q}$ is the spin angular momentum operator of the $q$th spin.

The complex mtdm vector $\muvec$ is the matrix element of the magnetic dipole moment operator for the transition, $\muvec=\langle b|\widehat{\muvec}|a\rangle$. $\muvec$ depends on the type of transition. In the simplest case, for a single-quantum transition in an isotropic spin system, $\muvec$ is complex and perpendicular to $\vect{B}_0$ and describes a rotation around the external magnetic field. For a zero-quantum transition in an isotropic system, $\muvec$ is real and parallel to $\vect{B}_0$. $\muvec$ is unique within an arbitrary complex phase factor.

A few vectors are needed to describe the experimental excitation geometry (Fig. \ref{fig:ex_shemes}). 
In EPR experiments usually a static magnetic field $\vect{B}_0$ is applied, hence its direction ${\vect{n}_0=\vect{B}_0/|\vect{B}_0|}$, is needed. For resonator and beam EPR experiments using linear polarization, $\vect{n}_1$ is required. The two limiting cases are perpendicular ($\vect{n}_1\!\perp\!\vect{n}_0$) and parallel ($\vect{n}_1\!\|\vect{n}_0$) mode. For beam experiments with unpolarized and circularly polarized radiation, the propagation direction ${\vect{n}_k=\vect{k}/|\vect{k}|}$, with the wave vector $\vect{k}$, is required. Two limiting cases are the Voigt geometry with ${\vect{n}_k\!\perp\!\vect{n}_0}$, and the Faraday geometry with ${\vect{n}_k||\vect{n}_0}$ (Fig. \ref{fig:ex_shemes}(b) and \ref{fig:ex_shemes}(c)).

Together with $\muvec$, the excitation geometry and the polarization determine the form of $D$ in Eq.~(\ref{eq:fermi}). For aligned spin centers in a single orientation, it is
\begin{subequations}\label{eq:general}
\begin{eqnarray}
D_\text{lin}^\text{s} & = &
\normsquared{\vect{n}_1^\text{T}\muvec},
\label{eq:D_lin}
\\
D_\text{un}^\text{s} & = &
\frac{1}{2}\left(\normsquared{\muvec}-\normsquared{\vect{n}_k^\text{T}\muvec}\right),
\label{eq:D_unpol}
\\
D_\pm^\text{s} & = &
2 D_\text{un}^\text{s}
\pm
2\vect{n}_k^\text{T}\!\left(\mr{Im}\muvec\!\times\!\mr{Re}\muvec\right),
\label{eq:D_circ}
\end{eqnarray}
\end{subequations}
for linear polarized ($D_\text{lin}^\text{s}$), unpolarized ($D_\text{un}^\text{s}$), and circular polarized light ($D_\pm^\text{s}$). $D_\text{lin}^\text{s}$ contains the projection of $\muvec$ onto $\vect{n}_1$, whereas $D_\text{un}^\text{s}$ contains the projection of $\muvec$ onto the plane perpendicular to $\vect{n}_k$. The $\pm$ in $D_\pm^\text{s}$ denotes right- and left-handed circular polarization, respectively. 
$D_\pm^\text{s}$ contains the unpolarized expression and an additional cross product term. If $\muvec$ is complex, the cross product represents the rotation axis for the rotation described by $\muvec$. If this rotation axis is perpendicular to $\vect{n}_k$, then left- and right-hand polarization give identical intensities. The cross product is independent of the overall phase of $\muvec$.
 
For disordered systems such as powders and glasses, the above expressions can be integrated. The resonance positions (frequencies and fields) of a spin center are generally anisotropic and are determined by the center's orientation relative to $\vect{n}_0$. Centers with identical orientation relative to $\vect{n}_0$, but different orientation relative to $\vect{n}_1$, have identical resonance positions, but differ in their line intensities. The integrals of the general expressions from Eq. (\ref{eq:general}) over this subset of spin centers give
\begin{subequations}\label{eq:general_powder}
\begin{eqnarray}
D_\text{lin}^\text{pow} & = &
\frac{1}{2}
\left[
\left(1-\xi_1^2\right)
\normsquared{\muvec}
+
\left(3\xi_1^2-1\right)
\normsquared{\vect{n}_0^\text{T}\muvec}
\right],
\label{eq:D_lin_pow}
\\
D_\text{un}^\text{pow} & = &
\frac{1}{4}
\left[
\left(1+\xi_k^2\right)
\normsquared{\muvec}
-
\left(3\xi_k^2-1\right)
\normsquared{\vect{n}_0^\text{T}\muvec}
\right],
\label{eq:D_unpol_pow}
\\
D_\pm^\text{pow} & = &
2D_\text{un}^\text{pow}
\pm
2\xi_k \vect{n}_0^\text{T}
\left(\mr{Im}\muvec\!\times\!\mr{Re}\muvec\right),
\label{eq:D_circ_pow}
\end{eqnarray}
\end{subequations}
with ${\xi_1 = \vect{n}_1^\text{T}\vect{n}_0}$ and ${\xi_k = \vect{n}_k^\text{T}\vect{n}_0}$. These expressions simplify for several common cases. For linear polarization, perpendicular (${\vect{n}_1\!\perp\! \vect{n}_0}$) and parallel (${\vect{n}_1 || \vect{n}_0}$) modes give
\begin{equation*}
D_{\text{lin,}\perp}^\text{pow} =\frac{1}{2}\left(\left|\muvec\right|^2-\left|\vect{n}_0^\text{T}\muvec\right|^2\right)\quad\text{and}\quad
D_{\text{lin,}||}^\text{pow} = \left|\vect{n}_0^\text{T}\muvec\right|^2.
\end{equation*}
Since both expressions cannot be zero simultaneously unless $\muvec$ is zero, we see that any transition with a finite mtdm will give intensity in at least one of the two modes.

For unpolarized radiation we can distiguish between ${\vect{n}_k\!\perp\!\vect{n}_0}$ (Voigt geometry) (see also \cite{Lund04, Lund06}), and ${\vect{n}_k||\vect{n}_0}$ (Faraday geometry), where we obtain
\begin{equation*}
D_{\text{un,V}}^\text{pow} = \frac{1}{4}\left(\left|\muvec\right|^2+\left|\vect{n}_0^\text{T}\muvec\right|^2\right),
D_\text{un,F}^\text{pow} =  \frac{1}{2}\left(\left|\muvec\right|^2-\left|\vect{n}_0^\text{T}\muvec\right|^2\right).
\end{equation*}
$D_\text{un,F}^\text{pow}$ is identical to $D_{\text{lin,}\perp}^\text{pow}$ and $D_\text{un,F}^\text{s}$. For Voigt geometry, ${D_{+}^\text{pow}=D_{-}^\text{pow}}$ as a result of the powder integration, therefore right- and left-handed circular polarization give the same line intensities, independent of the particular transition or the internal structure of the spin center.

For the case of an isotropic spin system, Eqs. (\ref{eq:general}) and Eqs. (\ref{eq:general_powder}) are equivalent. In this case, $\muvec$ is either perpendicular or parallel to $\vect{n}_0$.

For isolated electron spins in the absence of any interaction except the electron Zeeman interaction, the mtdm for an allowed transition ${|m_S\rangle\rightarrow|m_S+1\rangle}$ can be calculated analytically by using a frame ${(\vect{i}, \vect{j}, \vect{u})}$ where $\vect{u}$ is the quantization axis,  $\vect{u}=\gmatrix^\text{T}\vect{n}_0/g$ with $g = \left|\gmatrix^\text{T}\vect{n}_0\right|$.
The frame is obtained by the Bleaney-Bir transformation \cite{Bleaney51, Schweiger76}. In this frame, the mdtm is given, within an arbitary phase factor, by $\muvec = c \gmatrix (\vect{i}-\ii\vect{j})$, with ${c = -\mu_\mathrm{B} \sqrt{S(S+1)-m_S(m_S+1)}/2}$.
For a single orientation,  we find \footnote{Several authors \cite{Kneubuehl61,Iwasaki74, Rockenbauer74} use the reverse order of terms in the electron Zeeman Hamiltonian. Their g-matrix is the transpose of our g-matrix.}:
\begin{subequations}\label{eq:isolated}
  \begin{eqnarray}
    \tilde{D}_\text{lin}^\text{s} & = &  c^2 \varLambda(\vect{n}_1) = c^2 \varGamma(\vect{n}_1),\label{eq:D_lin_s}\\
   \tilde{D}_\text{un}^\text{s} & = & \frac{1}{2}c^2\left[ \text{tr}(\gmatrix\gmatrix^\text{T})-|\gmatrix\vect{u}|^2-
    \varGamma(\vect{n}_k) \right],\label{eq:D_unpol_s}\\
    \tilde{D}_\pm^\text{s} & = & 2\tilde{D}_\text{un}^\text{s} \pm 2\,c^2 \xi_k\det(\gmatrix)/g,\label{eq:D_circ_s}
  \end{eqnarray}
\end{subequations}
with
${\varGamma(\vect{v}) = \normsquared{ \det(\gmatrix)\gmatrix^{-1}\left(\vect{v}\times\vect{n}_0\right)/g}}$
and
${\varLambda(\vect{v}) = \normsquared{\left(\gmatrix^\text{T}\vect{n}_1\right)\times\vect{u}}}$. These expressions depend only on $\vect{u}$, and not on $\vect{i}$ or $\vect{j}$. (The diacritic tilde is used to distinguish the expressions for this special case from the general ones.) 
$\tilde{D}_\text{lin}^\text{s} = c^2 \Lambda(\vect{n}_1)$ was first derived by Kneub{\"u}hl \cite{Kneubuehl61}. $\tilde{D}_\text{lin}^\text{s}$ expands to other previously published expressions \cite{Iwasaki74, Rockenbauer74, Lund08} and simplifies for diagonal $\gmatrix$ \cite{Pilbrow69, Isomoto70}. Bleaney's original expression \cite{Bleaney60} is a very special case where $\vect{n}_1$ is limited to a symmetry plane of the eigenframe of an axial g tensor.  $\tilde{D}_\pm^\text{s}$  was previously derived for Faraday geometry \cite{Rockenbauer76}.

For linear polarization, the dependence of the line intensity on the relative orientation between $\vect{n}_1$ and $\vect{n}_0$ is evident from $\varGamma(\vect{n}_1)$: maximal for ${\vect{n}_1\!\perp\!\vect{n}_0}$ and zero for ${\vect{n}_1||\vect{n}_0}$.
In the special case, the rotation axis of the rotation described by $\muvec$ is along $\vect{n}_0$ (without restrictions on $\gmatrix$), hence in Voigt geometry  ${\tilde{D}_{+}^\text{s}=\tilde{D}_{-}^\text{s}}$. 
Both left- and right-handed circular polarization give the same transition intensity. In the limiting case of an isotropic g matrix  (${\gmatrix=g_\mathrm{iso} \mathbf{1}}$) in Faraday geometry, we get ${\tilde{D}_{\pm\text{,F}}^\text{s}=2c^2 g_\mathrm{iso}^2 (1\pm 1)}$, and only one handedness leads to non-zero intensity. 

For powders of isolated spins, $\tilde{D}$ is
\begin{subequations}\label{eq:isolated_powder}
\begin{eqnarray}
\tilde{D}_\text{lin}^\text{pow} & = &
\frac{1}{2}c^2
\left(1-\xi_1^2\right)
\left[
\mathrm{tr}(\ggT) - \normsquared{\gmatrix\vect{u}}
\right],
\label{eq:D_lin_s_pow}
\\
\tilde{D}_\text{un}^\text{pow} & = &
\frac{1}{4}c^2
\left(1+\xi_k^2\right)
\left[
\mathrm{tr}(\ggT) - \normsquared{\gmatrix\vect{u}}
\right],
\label{eq:D_unpol_s_pow}
\\
\tilde{D}_\pm^\text{pow} & = &
2 \tilde{D}_\text{un}^\text{pow}
\pm
2\,c^2 \xi_k^2 \det(\gmatrix)/g.
\label{eq:D_circ_s_pow}
\end{eqnarray}
\end{subequations}
Special cases of $\tilde{D}_\text{lin}^\text{pow}$ and ${\vect{n}_1\perp\vect{n}_0}$ are known for general and diagonal g tensors \cite{Isomoto70,Rockenbauer74,Aasa75, Lund08} and spelled out into polar coordinates \cite{Bleaney60, Holuj66, Pilbrow69, vanVeen78}.


Eqs.~(\ref{eq:isolated}) and (\ref{eq:isolated_powder}) are fully analytical and do not require numerical matrix diagonalization. They apply only in the absence of fine interactions and interactions with other spins or nuclei. More complicated spin systems are described by the more general expressions from Eqs. (\ref{eq:general}) and (\ref{eq:general_powder}). These require matrix diagonalization to compute $|a\rangle$ and $|b\rangle$, needed to calculate $\vect{\mu}$. Although it is customary to define a $z$ axis along $\vect{B}_0$ and an $x$ axis along $\vect{B}_1$ (for linear polarized mw radiation with $\vect{B}_1\!\perp\!\vect{B}_0$), no such axis definitions are used here. All expressions are representation independent. The new expressions are implemented in the EPR simulation software \textit{EasySpin} \cite{Stoll06}.

\begin{figure}
  \includegraphics{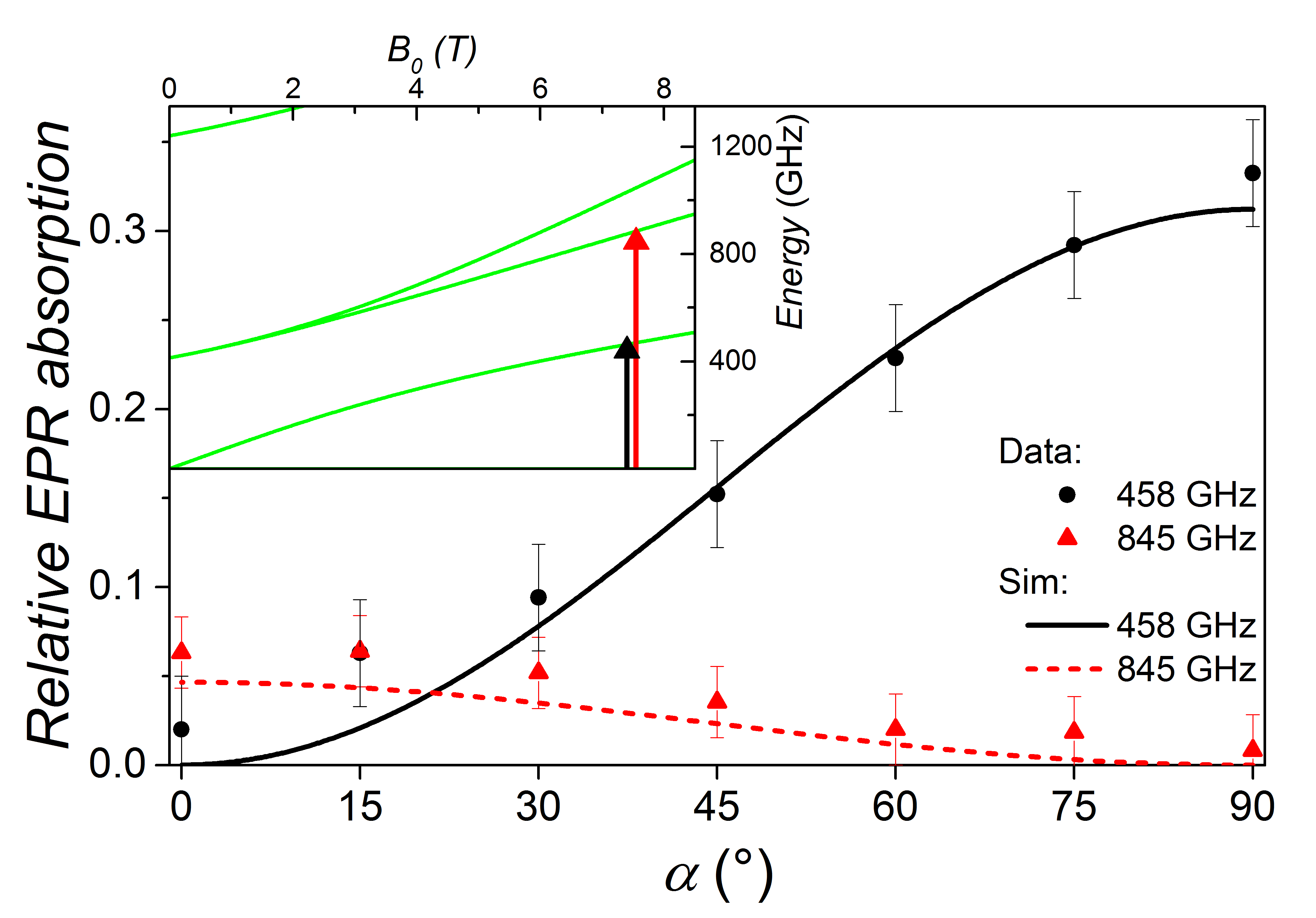}%
  \caption{Relative EPR absorption of high-spin Fe$^\text{III}$ (S = 5/2) in hemin at 458 GHz (black circles) and 845 GHz (red triangles) plotted vs. the polarization angle $\alpha$, which was in the used excitation geometry the angle between $\vect{n}_0$ and $\vect{n}_1$. Simulated relative EPR absorptions are plotted as black solid (458 GHz) and red dashed (845 GHz) lines, respectively. The inset depicts the calculated spin energies normalized to the ground state level as function of the magnetic field applied perpendicular to the hard axis of hemin. The black and red arrows mark the transitions observed at 458 GHz and 845 GHz, respectively.
  \label{fig:experiment}}
\end{figure}

Next, we validate the new theory experimentally and show that it can be used to extract new information about quantum systems from EPR data.  In a beam EPR experiment, we measured transition intensities of high-spin Fe$^\text{III}$ ions (spin $5/2$) in a powder sample of hemin as a function of the polarization angle $\alpha$ between the static magnetic field and the magnetic field component of the radiation. Experiments were conducted at the Frequency Domain Fourier Transform THz-EPR setup at BESSY II \cite{Schnegg09}. The high frequency/high field EPR data were obtained in Voigt geometry with a beam of broad-band unpolarized radiation that excited ground-state EPR transitions  at 7.5 T and 2 K in the frequency range between 400 GHz and 1200 GHz. Detection was achieved with a broad-band superfluid He cooled bolometer equipped with a wire grid in front of the detector. Further experimental details are given in the SM. During the experiment, the wire grid, which served as polarization selector, was rotated in steps from ${\vect{n}_1||\vect{n}_0}$ (${\alpha= 0^\circ}$) to ${\vect{n}_1\!\perp\!\vect{n}_0}$ (${\alpha = 90^\circ}$).

EPR transitions were observed at 458 GHz and 845 GHz. The two lines result from hemins oriented such that the external magnetic field is perpendicular to the molecular hard axis. Other orientations give rise to lines of much smaller intensity at other frequencies. As shown in Fig. \ref{fig:experiment}, the two lines have opposite polarization dependence: The line intensity at 458 GHz increases with increasing $\alpha$ while that of the 845 GHz line decreases. Fig. \ref{fig:experiment} shows theoretical intensities based on Eq. (\ref{eq:D_lin_pow}). They fit the experimental data well.

The EPR line at 458 GHz corresponds to a transition from the ground state to the first excited state (see inset in Fig. \ref{fig:experiment}), with opposite parity \footnote{The parity operator along $x$ is defined as ${\widehat{P}=\sum_m|m\rangle\langle-m|}$,  with the Zeeman state ${|m\rangle}$ along $z$. A state ${|\psi\rangle}$ with ${\langle\psi|\hat{P}|\psi\rangle = \pm1}$  is even for the $+$ sign, and odd for $-$ sign.}. For this transition, the mtdm is complex, perpendicular to $\vect{n}_0$ and has odd parity. The line intensity is proportional to ${(1-\xi_1^2)|\muvec|^2}=\normsquared{\muvec}\sin^2\!\alpha$ (see Eq.(\ref{eq:D_lin_pow})). On the other hand, the 845 GHz line corresponds to a transition from the ground state to the second excited state, which has the same parity as the ground state. $\muvec$ is real, parallel to $\vect{n}_0$, and of even parity. The transition rate is proportional to ${\xi_1^2 |\muvec|^2}=|\muvec|^2\cos^2\!\alpha$. The polarization dependence is reversed.

This EPR linear dichroism experiment shows that the polarization dependence of EPR line intensities can be used to determine parities of the states involved in EPR transitions. This is valuable structural information especially for more complicated spin centers like molecular nanomagnets, where the parity determines quantum tunneling rates \cite{Wernsdorfer99, Wernsdorfer02,Waldmann09, Henderson09}.
The experiment presented here was performed on a unique very-high-frequency EPR setup. However, the general theory outlined here might inspire similar experiments at lower frequencies, exploring easily rotatable microresonators \cite{Schoeppner14}, flexible and twistable waveguides, rotatable magnets, or vector magnets.

In conclusion, we derived novel general and compact closed-form expressions for calculating magnetic transition dipole moments and transition rates in solid-state EPR experiments on crystals and disordered materials. The expressions are valid for arbitrary spin centers and arbitrary excitation geometries. They cover resonator setups and beam experiments with unpolarized excitation, linear polarization and circular polarization. Furthermore, they are independent of the choice of the reference frame and do not involve Euler angles. Specification of the vectors $\vect{n}_0$, $\vect{n}_1$ or $\vect{n}_k$, and $\boldsymbol\mu$ is sufficient. The derivations show that the concept of magnetic transition dipole moment (both the operator and its transition matrix element) is useful for the description of general EPR experiments. With this new theory in hand, EPR spectra from experimental setups with non-standard geometries will give access to previously inaccessible information contained in line intensities and line shapes. This will have impact on high-field EPR and frequency-domain EPR, as well as on EPR experiments involving resonant microstructures with inhomogeneous fields and spatially varying excitation geometries.

\begin{acknowledgments}
This work has been supported and funded by Deutsche Forschungsgemeinschaft (DFG) priority program SPP 1601 and the University of Washington. The authors thank Robert Bittl (FU Berlin) for permitting measurements at the Frequency Domain Fourier Transform THz-EPR setup and Dirk Ponwitz (HZB) for technical support.
\end{acknowledgments}
\bibliography{polarization21}

\end{document}